# Observation of a nodal chain with Dirac surface states in TiB$_2$

C.-J. Yi,[1,2,†] B. Q. Lv,[1,2,†] Q. S. Wu,[3,4,†] B.-B. Fu,[1,2,†] X. Gao,[1,2] M. Yang,[1,2] X.-L. Peng,[1,2] M. Li,[5] Y.-B. Huang,[5] P. Richard[1,2,7] M. Shi,[6] G. Li,[1] Oleg V. Yazyev,[3,4] Y.-G. Shi,[1*] T. Qian,[1,2,7*] and H. Ding[1,2,7*]

[1] *Beijing National Laboratory for Condensed Matter Physics and Institute of Physics, Chinese Academy of Sciences, Beijing 100190, China*

[2] *University of Chinese Academy of Sciences, Beijing 100049, China*

[3] *Institute of Physics, École Polytechnique Fédérale de Lausanne (EPFL), CH-1015 Lausanne, Switzerland*

[4] *National Centre for Computational Design and Discovery of Novel Materials MARVEL, Ecole Polytechnique Fédérale de Lausanne (EPFL), CH-1015 Lausanne, Switzerland*

[5] *Shanghai Synchrotron Radiation Facility, Shanghai Institute of Applied Physics, Chinese Academy of Sciences, Shanghai 201204, China*

[6] *Paul Scherrer Institute, Swiss Light Source, CH-5232 Villigen PSI, Switzerland*

[7] *Collaborative Innovation Center of Quantum Matter, Beijing, China*

[†] These authors contributed to this work equally.

[*] Corresponding authors: ygshi@iphy.ac.cn, tqian@iphy.ac.cn, dingh@iphy.ac.cn

**Abstract**

Topological nodal-line semimetals (TNLSMs) are characterized by symmetry-protected band crossings extending along one-dimensional lines in momentum space. The nodal lines exhibit a variety of possible configurations, such as nodal ring, nodal link, nodal chain, and nodal knot. Here, using angle-resolved photoemission spectroscopy, we observe nodal rings in the orthogonal $k_z = 0$ and $k_x = 0$ planes of the Brillouin zone in $TiB_2$. The nodal rings connect with each other on the intersecting line Γ-K of the orthogonal planes, forming a remarkable nodal-chain structure. Furthermore, we observe surface states (SSs) on the (001) cleaved surface, which are consistent with the calculated SSs considering the contribution from both Ti and B terminations. The calculated SSs have novel Dirac-cone-like band structures, which are distinct from the usual drumhead SSs with a single flat band proposed in other TNLSMs.

Topological semimetals (TSMs) are characterized by symmetry-protected band degeneracy in their electronic structure near the Fermi level ($E_F$) [1-12]. The TSMs can be divided into two groups according to the dimensionality of band degeneracy in momentum space. The first group has zero-dimensional (0D) band degeneracy, e.g. the Dirac and Weyl semimetals, in which two doubly- or singly-degenerate bands cross each other, forming the 4-fold Dirac points or 2-fold Weyl points [1-11]. The fermionic excitations near Dirac and Weyl points are analogous to Dirac and Weyl fermions in high-energy physics, respectively. Moreover, band theory has suggested several other types of TSMs with 3-, 6-, and 8-fold degenerate nodal points [13-17], in which the fermionic excitations have no analogues in high-energy physics. The TSMs with 3-fold points have been experimentally confirmed in MoP and WC recently [18,19].

The second group has 4- or 2-fold band crossing along one-dimensional (1D) lines in momentum space. The materials with such nodal lines are called topological nodal-line semimetals (TNLSMs) [11,12]. In contrast to the 0D nodal points, the 1D nodal lines form a much larger variety of possible configurations, such as nodal rings, nodal links, nodal chains, and nodal knots [Fig. 1(a)] [20-28]. Theoretical calculations have shown that the different types of nodal lines have distinct Berry phases and Landau levels, which could induce different magnetotransport and magneto-optical properties [24,28]. As compared with 1D Fermi arc surface states (SSs) in the TSMs with 0D nodal points, the TNLSMs are proposed to host two-dimensional (2D) drumhead SSs [29]. Among the TNLSM candidates, several materials have been studied by angle-resolved photoemission spectroscopy (ARPES), which provides experimental evidence of extended nodal lines and closed nodal rings [30-32]. In addition, a recent study has discovered a nodal-chain structure in a photonic crystal [33].

In this work, we use ARPES to systematically investigate the electronic structures of both bulk and surface states of $TiB_2$. Our results clearly reveal that the nodal rings in the vertical and horizontal mirror planes connect with each other, unambiguously demonstrating the nodal-chain structure in $TiB_2$. Furthermore, the observed SSs on the (001) surface are remarkably well reproduced by our first-principles calculations, which show novel Dirac-cone-like band structures distinct from the usual drumhead SSs in other TNLSMs.

High-quality single crystals of TiB$_2$ were grown by the flux method. ARPES measurements were performed at the "Dreamline" beamline of the Shanghai Synchrotron Radiation Facility (SSRF) and at the Surface and Interface Spectroscopy (SIS) beamline at Swiss Light Source (SLS). The samples were cleaved *in situ* and measured at 20 K in a vacuum better than $5 \times 10^{-11}$ Torr. The DFT calculations were performed using the projector augmented-wave method [34] and the PBE-GGA exchange-correlation functional [35] as implemented in VASP code [36]. The slab spectrum functions were calculated by using Wannier Tools with the maximally localized Wannier functions [37,38] based tight-binding model.

TiB$_2$ has a layered structure with space group *P*6/*mmm* (No. 191), which consists of alternating stacking of close-packed hexagonal Ti layers and graphene-like honeycomb lattice B layers, as illustrated in Fig. 1(b). The calculations have predicted multiple nodal lines [21], whose momentum locations in the BZ are illustrated in Fig. 1(d). These nodal lines are protected by $\mathcal{PT}$ symmetry, while the mirror symmetries $\sigma_h$, $\sigma_{v1}$, $\sigma_{v2}$ and $\sigma_{v3}$ in the $D_{6h}$ group enforce them embed on mirror planes. They can be divided into four classes, including the nodal ring surrounding K point in the horizontal Γ-K-M ($k_z = 0$) plane (class A), the nodal ring in the vertical Γ-K-H-A ($k_x = 0$) plane (class B), the nodal line along Γ-A starting from a triple point (class C), and the nodal ring surrounding A point in the horizontal A-H-L ($k_z = \pi$) plane (class D). Remarkably, the class-A and class-B nodal rings connect with each other at one point on Γ-K line, which is the intersecting line of the $k_z = 0$ and $k_x = 0$ planes, forming a nodal-chain structure.

Figure 1(e) shows the calculated band structure on several high-symmetry lines in the absence of SOC. Among them, the Dirac-like band crossings on Γ-H, Γ-K and K-M lines are clearly identified. The band crossings along Γ-K and K-M form the class-A nodal ring, while those along Γ-K and Γ-H form the class-B nodal ring. Note that the class-A and class-B nodal rings share the same band crossing along Γ-K, resulting in the nodal-chain structure. When SOC is considered, the crossing points open small gaps of ~ 20 meV, which is common in $\mathcal{PT}$ protected systems. The millivolt level gaps indicate that the effect of SOC on the electronic structure of TiB$_2$ is quite weak and thus we ignore the gaps in our experimental data.

We first use ARPES to investigate the electronic structures in the vertical mirror planes. All the ARPES measurements were performed on the (001) cleaved surface. By varying incident photon energy (*hv*), we obtained the FSs in the two vertical planes Γ-M-L-A ($k_y = 0$) and Γ-K-H-A ($k_x = 0$), as shown in Figs. 2(a) and 2(b), respectively. The FSs exhibit a modulation along $k_z$ with a period of 2π/*c*, confirming their bulk character. The results show a nearly rectangular FS surrounding A point in the two planes. The most remarkable difference between the measured FSs in the two planes is the FS lines that connect to the rectangular FSs in the $k_x = 0$ plane. In addition, we observe some FS lines along $k_z$ outside the rectangular FSs, which should be attributed to the SSs. The measured bulk FSs are consistent with the calculations in Figs. 2(c) and 2(d). Figure 2(f) shows that one band disperses towards $E_F$ upon moving from Γ to K, in agreement with the band calculations in Fig. 2(i), but it is hard to identify another Dirac band in the experimental data in Figs. 2(f) and 2(i). In the second BZ, we clearly observe the two Dirac bands along Γ-K in Figs. 2(g) and 2(j). The results in Fig. 2 confirm the nodal rings in the $k_x = 0$ plane.

To identify the nodal rings in the horizontal planes, we have investigated the electronic structures in the $k_z = 0$ and π planes. The results at $k_z = 0$ in Fig. 3(a) exhibit a triangular FS surrounding K point, while those at $k_z = π$ in Fig. 3(c) exhibit a point-like FS at A point surrounded by a circular FS. The measured FSs are consistent with the calculations in Figs. 3(b) and 3(d). The band crossings of the nodal ring at $k_z = π$ cannot be observed because these degeneracies lie above $E_F$. We mainly focus on the nodal ring in the $k_z = 0$ plane. Figures 3(e) and 3(g) show the measured band dispersions along Γ-K-M and K-M-K, respectively. The results clearly show a Dirac-like band crossing along both Γ-K and K-M. While the crossing point along Γ-K is very close to $E_F$, the one along K-M is located at ~ 0.5 eV below $E_F$, which is consistent with the calculations in Figs. 3(f) and 3(h). The results confirm the nodal ring configuration around K point in the $k_z = 0$ plane.

In Fig. 3(i), we combine the FSs measured in the $k_z = 0$ and $k_x = 0$ planes. The three-dimensional (3D) plot clearly shows that the class-A and class-B nodal rings connect with each other on Γ-K line, in good agreement with the calculated FSs in Fig. 3(j), unambiguously demonstrating the nodal-chain structure in $TiB_2$.

In addition to the Dirac nodal chain in the bulk electronic structures of $TiB_2$, we also observed SSs on the freshly cleaved (001) surface. Figures 4(a) and 4(e) show the experimental FSs and band dispersions, which are distinct from the calculated bulk electronic structures in any $k_x$-$k_y$ plane. Since the cleavage takes place between the adjacent Ti and B layers, two kinds of surface terminations, *i.e.* the Ti and B terminations, are possible. We have calculated the FSs and band dispersions of the SSs for a (001) slab system with the Ti and B terminations, respectively, as shown in Figs. 4(b), 4(c), 4(f), and 4(g). Assuming that both surface terminations are present in our samples, we combine together the calculated results for the two terminations in Figs. 4(d) and 4(h). The experimental results are in remarkably good agreement with the combined plots. By comparing with the calculated SSs, we assign the extracted experimental band dispersions in Fig. 4(i) to the SS bands of the Ti and B terminations, respectively.

The calculations in Figs. 4(f) and 4(g) show that the SSs have Dirac-cone-like band structures around $\bar{K}$ point on both Ti and B terminations, in sharp contrast to the drumhead SSs with a single flat band proposed in other TNLSMs [29]. In usual nodal lines, the band inversion occurs between two involved bands, inducing drumhead SSs, as shown in Fig. 4(l). In $TiB_2$ there are four bands involved in the nodal ring around the $K$ point, where the band inversion occurs twice, inducing two drumhead surface states [22]. The results in Figs. 3(e)-3(h) show that two bulk bands associated with the nodal ring cross each other at -2.0 eV at K point, confirming the four-band nodal ring in $TiB_2$. The two SSs are doubly degenerate at the $\bar{K}$ point, due to protection of the $C_{3v}$ symmetry, forming the Dirac-cone-like band structures, as shown in Fig. 4(k).

The Dirac SSs in $TiB_2$ are also distinct from those for topological insulators since the $\bar{K}$ point is not invariant under time-reversal operation. The calculations indicate that the Dirac SSs on B and Ti terminations have linear and quadratic band dispersions, which are analogous to those of monolayer and bilayer graphene, respectively [21]. The Berry phases around the two kinds of Dirac cones are π and 2π, respectively, which would lead to different quantum oscillations. As the calculated Dirac points for both Ti and B-terminated surfaces lie above $E_F$, they cannot be detected in the ARPES experiments. However, the excellent consistency between experiment and calculation provides a compelling evidence for the existence of surface Dirac cones.

Up to now, the nodal-line, nodal-ring, and nodal-chain materials have been theoretically predicted and experimentally demonstrated, whereas the materials with nodal-link and nodal-knot structures have not been realized. While theoretical considerations have proposed several symmetry protection mechanisms for the TNLSMs, the classification of TNLSMs is yet to be complete. Recent transport measurements on TNLSMs under high magnetic fields have revealed exotic properties [39,40], which were attributed to the nodal-line structures. Further theoretical and experimental efforts are needed to better understanding and applications of TNLSMs.

# Acknowledgements

We thank Mengyu Yao, Junzhang Ma, and M.Naamneh for the assistance in the ARPES experiments. We also thank Chen Fang and Hongming Weng for the discussion. This work was supported by the Ministry of Science and Technology of China (2016YFA0300600, 2016YFA0401000, and 2017YFA0302901), the National Natural Science Foundation of China (11622435, 11474340, 11474330, and 11774399), the Chinese Academy of Sciences (XDB07000000, XDPB08-1, and QYZDB-SSW-SLH043), and the Beijing Municipal Science and Technology Commission (No. Z171100002017018). Q.W., M.S. and O.V.Y. acknowledge support by the NCCR Marvel. First-principles electronic structure calculations have been performed at the Swiss National Supercomputing Centre (CSCS) under Project No. s832. Y.B.H. acknowledges support by the CAS Pioneer "Hundred Talents Program" (type C).

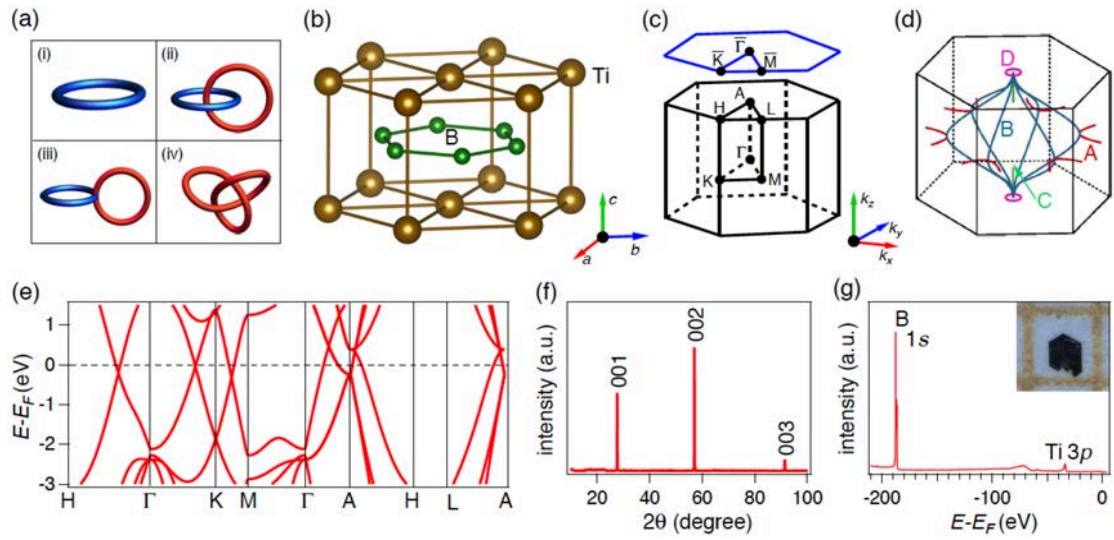

FIG. 1. (a) Schematic illustrations of four topological configurations formed by nodal lines: (i) nodal ring, (ii) nodal link, (iii) nodal chain, and (iv) nodal knot. (b) Crystal structure of $TiB_2$. (c) 3D bulk BZ and projected (001) surface BZ of $TiB_2$. $k_x$, $k_y$, and $k_z$ are along Γ-M, Γ-K, and Γ-A directions, respectively. (d) Momentum locations of the four classes of nodal lines in the BZ of $TiB_2$. (e) Calculated band structures of $TiB_2$ along several high-symmetry lines without SOC. (f) X-ray diffraction data measured on the (001) plane of $TiB_2$ single crystal. (g) Core level photoemission spectrum showing the characteristic Ti 3p and B 1s peaks. The inset shows a typical single crystal of $TiB_2$.

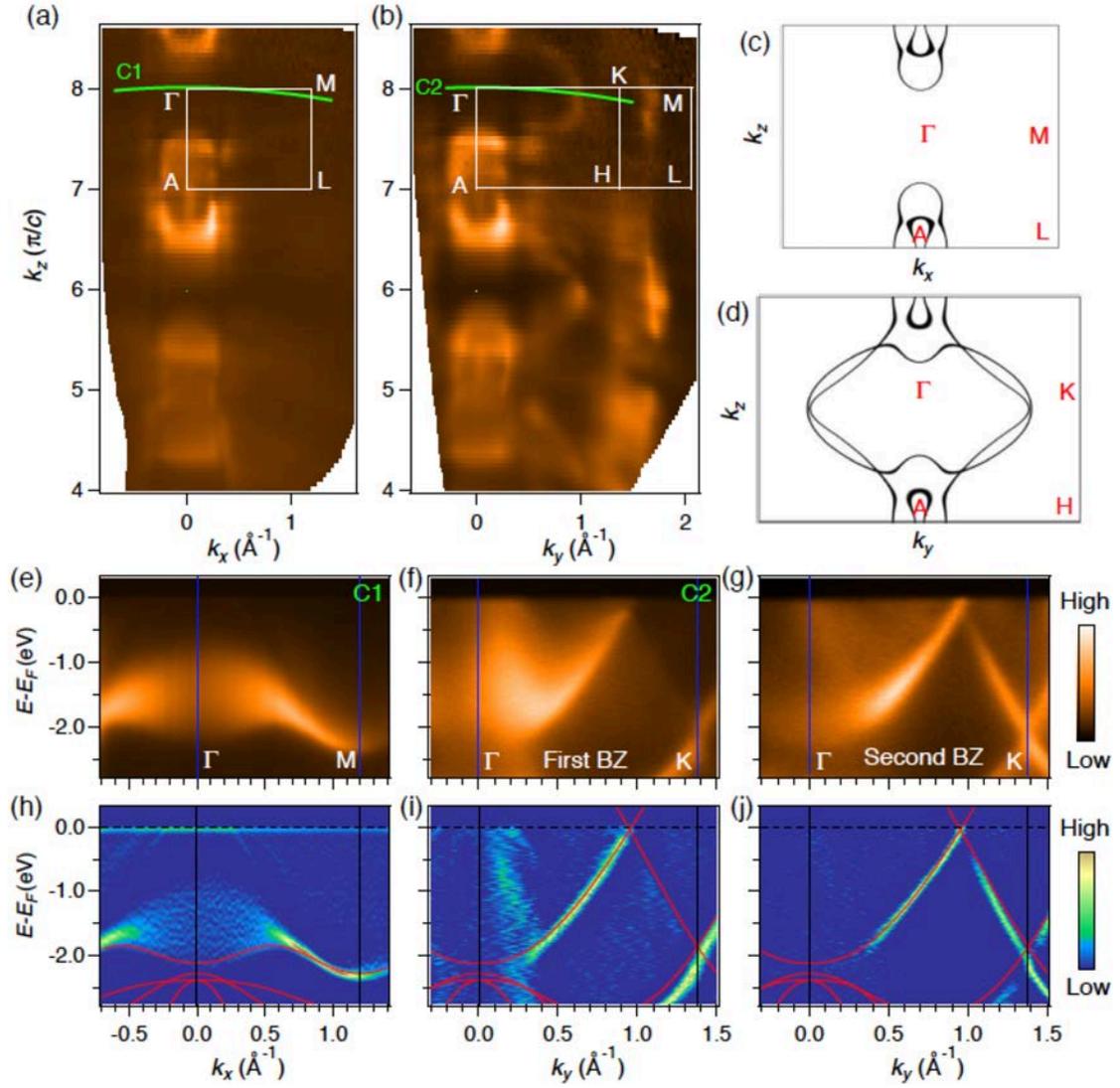

FIG. 2. (a,b) Experimental FSs in the Γ-M-L-A ($k_y = 0$) and Γ-K-H-A ($k_x = 0$) planes, respectively. (c,d) Corresponding calculated FSs. (e,f) ARPES intensity plots showing experimental band dispersions along Γ-M and Γ-K, respectively. (g) Same as (f) but measured in the second BZ. (h-j) Corresponding curvature intensity plots. For comparison, the calculated bands are superposed on top of the experimental data.

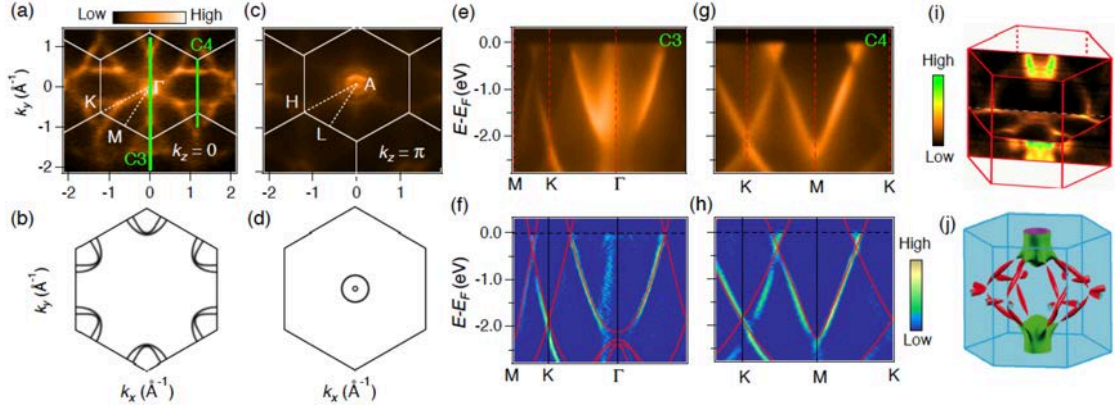

FIG. 3. (a,b) Experimental and calculated FSs in the $k_z = 0$ plane, respectively. (c,d) Experimental and calculated FSs in the $k_z = \pi$ plane, respectively. (e,f) ARPES and curvature intensity plots, respectively, showing band dispersions along Γ-K-M, indicated as C3 in (a). (g),(h) ARPES and curvature intensity plots, respectively, showing band dispersions along K-M-K, indicated as C4 in (a). For comparison, the calculated bands are superposed on top of the experimental data in (f,h). (i) 3D plot of the combined experimental FSs in the $k_z = 0$ and $k_x = 0$ planes. (j) Calculated 3D FSs in the BZ.

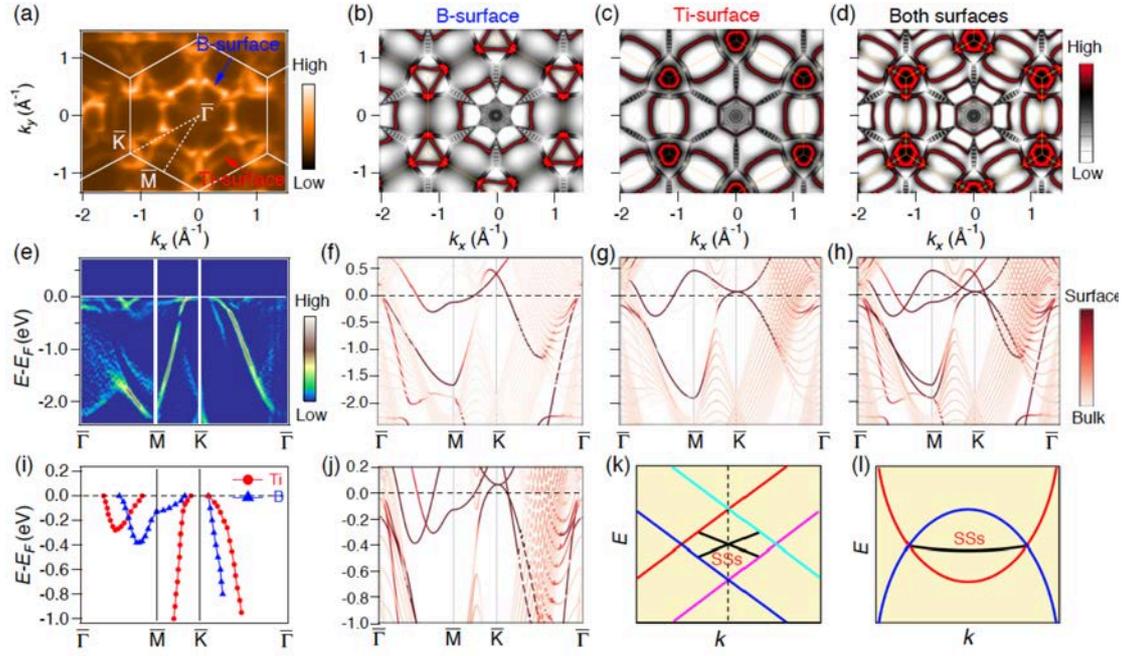

FIG. 4. (a) ARPES intensity plot at $E_F$ measured on a freshly cleaved (001) surface. (b,c) Calculated SS FSs for a (001) slab system with the B and Ti terminations, respectively. (d) Combination of the results in (b) and (c). (e) Curvature intensity plot showing band dispersions along the high-symmetry path $\bar{\Gamma}$-$\bar{M}$-$\bar{K}$-$\bar{\Gamma}$. (f,g) Calculated SS bands for a (001) slab system with the B and Ti terminations, respectively. (h) Combination of the results in (f) and (g). (i) Band dispersions extracted from the experimental data in (e). Red and blue symbols represent the SS bands of the Ti and B terminations, respectively. (j) Same as (h) but with the energy range same as in (i) for direct comparison. (k,l) Schematic illustrations of the four- and two-band nodal rings with Dirac and drumhead SSs, respectively.